\documentclass[fleqn,usenatbib]{mnras}

\usepackage{newtxtext,newtxmath}
\usepackage{caption}
\usepackage{tikz}

\usepackage[T1]{fontenc}

\DeclareRobustCommand{\VAN}[3]{#2}
\let\VANthebibliography\thebibliography
\def\thebibliography{\DeclareRobustCommand{\VAN}[3]{##3}\VANthebibliography}

\usepackage{graphicx}	
\usepackage{amsmath}	
\usepackage{subcaption}
\usepackage{tikz}
\usetikzlibrary{calc}

\newcommand{\Bv}{\boldsymbol{B}}

\newcommand{\uv}{\boldsymbol{u}}
\newcommand{\wv}{\boldsymbol{w}}

\newcommand{\jv}{\boldsymbol{j}}
\newcommand{\Rv}{\boldsymbol{R}}

\newcommand{\bv}{\boldsymbol{b}}
\newcommand{\Sigv}{\boldsymbol{\Sigma}}

\title[FL slippage rates in NLFFF extrapolations]{Field line slippage rate signatures in nonlinear force-free field extrapolations}

\author[S. Stanish et al.]{
Sage Stanish$^{1}$
and David MacTaggart$^{1}$
\\
$^{1}$School of Mathematics and Statistics, University of Glasgow, Glasgow, G12 8QQ, UK
}

\date{Accepted XXX. Received YYY; in original form ZZZ}

\pubyear{\the\year{}}

\begin{document}
\label{firstpage}
\pagerange{\pageref{firstpage}--\pageref{lastpage}}
\maketitle

\begin{abstract}
Magnetic reconnection plays a central role in solar flares and coronal mass ejections. Identifying where reconnection is physically active within coronal magnetic field models is a key part of magnetic field analysis. We investigate the field line slippage rate as a physics-weighted proxy for three-dimensional reconnection in nonlinear force-free field (NLFFF) extrapolations. The slippage rate measures the instantaneous deviation of magnetic field lines from ideal evolution, due to non‑ideal terms in Ohm’s law, providing a direct link between magnetic geometry and reconnection physics. For NLFFFs, we show that the resistivity-induced slippage rate is governed by cross-field gradients of the field-aligned twist, thus establishing a clear connection between current structure and reconnection signatures. We further examine its relationship to the squashing factor $Q$, used to identify quasi-separatrix layers (QSLs). By deriving a scaling estimate, we demonstrate that strong magnetic squashing amplifies slippage only insofar as it produces small transverse length scales; large values of $Q$ alone do not guarantee significant reconnection. We apply this framework to a sequence of NLFFF extrapolations of NOAA active region 11158 spanning the X2.2 flare of 15 February 2011. The slippage rate reveals enhanced reconnection signatures associated with distinct phases of the active region’s evolution. In comparison with the squashing factor, we show that the field line slippage rate provides a physics-weighted complement to QSL analysis, distinguishing between regions that are geometrically favourable for reconnection and those where reconnection is physically significant.

\end{abstract}

\begin{keywords}
magnetic reconnection -- Sun: magnetic fields -- Sun: corona
\end{keywords}



\section{Introduction}
Magnetic extrapolations from photospheric magnetograms form a key component of the observational analysis of solar active regions (ARs) and allow for the three-dimensional (3D) reconstruction of an AR's magnetic field. Various techniques have been developed based on different physical assumptions. In the corona, where the magnitude of the Lorentz force is small on large scales, the nonlinear force-free field (NLFFF) approximation has proven to be particularly effective \citep{WiegelmannSakurai2021} and there exists a variety of different methods for calculating such fields \citep[e.g.][]{Amari1999,Wheatland2000,Wiegelmann2004,YanLi2006,Valori2007,Inoue2014,Jarolim2023NatAstron}.

By construction, magnetic extrapolations are static (they are force balances) and so do not describe dynamics explicitly. However, that does not mean that they cannot provide useful information for important dynamical events, such as flares and coronal mass ejections (CMEs). The key strength of extrapolations is that they can reproduce many of the important geometrical and topological elements of an AR magnetic field. Such properties help to determine where magnetic reconnection (related to a particular flare or CME) is expected to occur and there are several approaches to determine such locations. One approach is to construct a \emph{magnetic skeleton}, which divides an AR magnetic field into distinct topological regions of connectivity \citep[e.g.][]{Priest1997,Longcope2005,Guo2019}. A prerequisite for this is the existence of null points in the AR extrapolation. The null points are connected by separator field lines and the resulting boundaries, known as \emph{separatrix surfaces}, are the expected locations for reconnection due to the fact that the field line connectivity undergoes discrete jumps in these locations. 

When null points are not present in a given magnetic field, which can happen in NLFFF extrapolations, a generalization of magnetic separatrix boundaries, known as \emph{quasi-separatrix layers} (QSLs), is required. QSLs were first introduced by \cite{PriestDemoulin1995} and then refined by  \cite{Titov1999QSL} and \cite{Titov2002Q}. They represent regions of strong but smooth connectivity gradients and are, therefore, geometrical features. However, since QSLs are also generalizations of separatrix boundaries which are topological, we refer to QSLs as quasi-topological boundaries. 

QSLs have played an important role in our understanding of the nature of large-scale reconnection \citep[e.g.][]{Aulanier2005,Titov2009SlipQ} in relation to flares and CMEs     \citep[e.g.][]{Demoulin1996,Janvier2013}. In  NLFFF extrapolations, QSLs provide a proxy for the probable locations of reconnection based purely on the geometry of the magnetic field and confirmation of their efficacy has been made through comparisons of QSL footpoints to flare ribbon locations \citep[e.g.][]{Masson2009,Savcheva2012QSL}.  

Other proxies related to reconnection that are based on magnetic field geometry, such as twist \citep[][]{BergerPrior2006}, are also used in conjunction with QSLs. While these quantities identify where reconnection is likely, they do not distinguish between regions that are merely geometrically favourable and those in which reconnection is physically active. Recently, the concept of the \emph{field line slippage rate} was introduced by \cite{MacTaggart2025Slippage}, following work in turbulence theory by \cite{Eyink2015TGMR}, as a means for identifying locations of reconnection. Like QSLs, the field line slippage rate (hereafter, \emph{slippage rate}) encodes the geometry of the magnetic field. However, in contrast to other reconnection proxies, it connects the magnetic field geometry to the non-ideal terms in Ohm's law. Thus, rather than being just indicative of possible reconnection, {if the slippage rate is non-zero, the magnetic field locally deviates from ideal evolution due to non-ideal terms in Ohm’s law, indicating the presence of connectivity-changing non-ideal effects.}

The use of the slippage rate as a proxy of reconnection in the analysis of NLFFF extrapolations was suggested by \cite{MacTaggart2025Slippage}. It is the purpose of this paper to explore this application in more detail and to show the effectiveness of the slippage rate in revealing more refined information about the nature of reconnection. To do this, we first {specify the basic conditions under which field line slippage occurs in relation to coronal applications}. We then describe the form of the slippage rate for a NLFFF and how it should be interpreted as a proxy of reconnection. This description is followed by a theoretical analysis of how the magnitude of the slippage rate relates to the squashing factor of QSLs. With the basic theory in place, we then proceed to analyze the behaviour of the slippage rate in sequences of NLFFF extrapolations of NOAA AR11158. In particular, we consider extrapolations in the build-up and immediate aftermath of an X2.2 flare on 15th February 2011. We further analyze the slippage rate in relation to flare ribbons and provide practical demonstrations of how slippage rate signatures relate to squashing factor signatures. The paper ends with a summary and a discussion.

\section{Basic conditions for field line conservation}\label{sec:reco_defns}

There exist many detailed works reviewing reconnection \citep[e.g.][]{LiPriestGuo2021,PontinPriest2022}. For completeness, however, we present some key concepts in order to clarify the meaning of the slippage rate and how it may then be used as a proxy for reconnection in NLFFF extrapolations.

{
In studies of reconnection in the solar corona, particularly on scales comparable to ARs, the plasma can be modelled as a single fluid. We, therefore, take as a starting point for our study of reconnection,  the non-ideal induction equation
\begin{equation}\label{induciton_non_ideal}
\frac{\partial \Bv}{\partial t}
= \nabla \times (\uv \times \Bv) - \nabla \times \Rv,
\end{equation}
where  $\Bv$ is the magnetic field, $\uv$ is a (smooth) bulk plasma velocity field and $\Rv$ represents non-ideal effects. The case $\Rv=\boldsymbol{0}$ corresponds to ideal magnetohydrodynamics (MHD), in which both magnetic flux and field lines are frozen into the plasma velocity.

Our primary interest here is to study reconnection through changes in field line connectivity with respect to ideal MHD. In order to address this, we must first consider under what general conditions field line connectivity is preserved. A long-established result, known as the \emph{Helmholtz-Zorawski criterion} \citep[][]{Truesdell1954Kinematics}, states (in relation to our application) that if field lines of $\Bv$ coincide with material lines of the plasma, with a transport velocity $\wv$, then
\begin{equation}\label{Helm_Zor}
\Bv\times\left[\frac{\partial\Bv}{\partial t}-\nabla\times(\wv\times\Bv)\right] = \boldsymbol{0}.
\end{equation}
For our application, we will only consider $\wv=\uv$ since we describe reconnection in terms of changes of field line connectivity relative to ideal plasma motion. For further discussion about more general choices of the transport velocity, the reader is guided to \cite{HornigSchindler1996} and \cite{Schindler2006SpacePlasma}. 

By comparing equations (\ref{induciton_non_ideal}) and (\ref{Helm_Zor}), it follows that the condition for field line conservation to hold is
\begin{equation}\label{fl_cons}
    \nabla\times\Rv = \lambda\Bv,
\end{equation}
for some scalar function $\lambda$. If equation (\ref{fl_cons}) is satisfied, the non-ideal term $\Rv$ preserves field line connectivity. Therefore, for genuine reconnection with a change in field line connectivity, equation (\ref{fl_cons}) must be violated (at least locally within the magnetic field). Equation (\ref{fl_cons}) is another way of expressing Newcomb's condition for field line conservation \citep[][]{Newcomb1958}.

In this work, we select a resistive non-ideal term
\begin{equation}
    \Rv=\eta\jv,
\end{equation}
where $\jv=\mu_0^{-1}\nabla\times\Bv$, with $\mu_0$ being the permeability of free space and $\eta$ the resistivity. In coronal applications, it is typical for $\eta$ to be modelled as a constant or as a function of current density. We may now write the induction equation as
\begin{equation}\label{resistive_induction}
    \frac{\partial \Bv}{\partial t}
= \nabla \times (\uv \times \Bv) - \nabla \times (\eta\mu_0^{-1}\nabla\times\Bv).
\end{equation}
Although the resistive term on the right-hand side of equation (\ref{resistive_induction}) may, in practice, be non-zero throughout the plasma, it is clear that it will be largest in regions where the magnetic field changes rapidly over short length scales, i.e. over (quasi-)topological boundaries. This term will induce a change in field line connectivity if it violates equation (\ref{fl_cons}), i.e. if it has a component perpendicular to the magnetic field. 
}

\section{Field line slippage rate}\label{slippage_rate_section}
As mentioned earlier, the (quasi-)topology of the magnetic field in the corona plays a fundamental role in determining the location of reconnection within an AR. This fact can be connected directly to a measure of reconnection through the induction equation (\ref{resistive_induction}). First we define the \emph{dissipation rate} as
\begin{equation}\label{SD}
    \Sigv_D = \frac{1}{|\Bv|}\left[\frac{\partial\Bv}{\partial t}-\nabla\times(\uv\times\Bv)\right] = -\frac{\nabla\times(\eta\mu_0^{-1}\nabla\times\Bv)}{|\Bv|}.
\end{equation}
This definition is simply the non-ideal part of the induction equation divided by the magnitude of the magnetic field, which results in $\Sigv_D$ being a rate. By definition, $\Sigv_D$ is only well-defined wherever the magnetic field strength is non-zero, which is suitable for describing reconnection away from null points.

\cite{AgarwalBhattacharyya2024} measure $|\Bv|\Sigv_D$ to find signatures of reconnection in simulations (to be precise, they measure the numerical dissipation that results from deviations from ideal MHD). It is clear that $\Sigv_D$ will be strong at (quasi-)boundaries and will thus provide a measure of magnetic dissipation in these regions. From equation (\ref{SD}), if $\Sigv_D=\boldsymbol{0}$ then the plasma motion is ideal. However, $\Sigv_D\ne\boldsymbol{0}$ does not necessarily imply reconnection, i.e. field line slippage induced by resistivity. As a simple example, consider the case of a linear force-free magnetic field with 
\begin{equation}\label{lfff}
    \uv=\boldsymbol{0} \quad {\rm and} \quad \nabla\times\Bv=\alpha_0\Bv
\end{equation}
where $\alpha_0$ is a constant. Assuming a constant $\eta$, we then have 
\begin{equation}\label{Sigma_D}
    \Sigv_D=-\eta\mu_0^{-1}\alpha_0^2\frac{\Bv}{|\Bv|}.
\end{equation}

{For the conditions given in equation (\ref{lfff}) together with a constant-$\eta$ resistivity, it is clear that the non-ideal term satisfies (\ref{fl_cons}).} Therefore, the dissipation term in equation (\ref{Sigma_D}), which is purely along the direction of the magnetic field, affects only the field strength without changing the field line topology \citep[see][for further discussion]{PriorMacTaggart2020}. In general, for more complex magnetic fields, $\Sigv_D$ will contain information about dissipation and field line slippage together.

In order to isolate the field line slippage of reconnection, we can consider the \emph{slippage rate} which, as shown by \cite{MacTaggart2025Slippage}, can be written as
\begin{equation}\label{Sigma}
    \Sigv = \frac{\partial\bv}{\partial t}-\frac{1}{|\Bv|}\left[\nabla\times(\uv\times\Bv)\right]_\perp = -\frac{[\nabla\times(\eta\mu_0^{-1}\nabla\times\Bv)]_\perp}{|\Bv|},
\end{equation}
where $\bv=\Bv/|\Bv|$ and a quantity enclosed by $[\cdot]_\perp$ means that the we consider its projection orthogonal to the direction of the magnetic field. {The slippage rate $\Sigv$ can be derived from the instantaneous motion of $\bv$ \citep[][]{MacTaggart2025Slippage}.} 

From equation (\ref{Sigma}), it is clear that if $\Sigv=\boldsymbol{0}$ then the field line motion is ideal {with respect to the plasma velocity $\uv$. Further, and unlike the case for $\Sigv_D$, $\Sigv\ne\boldsymbol{0}$  corresponds to the presence of a non-zero component of $\nabla\times\Rv$ perpendicular to $\Bv$ and, hence, to a local violation of the conditions for ideal field line evolution.}Therefore,  if $\Sigv\ne\boldsymbol{0}$ at a particular location at a particular time, the field line instantaneously at that location is undergoing resistivity-induced slippage. 

{We note a potential terminological ambiguity. In parts of the solar physics literature, the phrase “field line slippage” (or “slippage velocity”) refers to the velocity of a field line undergoing slippage. In contrast, the quantity $\Sigv$ is not a velocity but a rate of deviation of the magnetic field direction from ideal evolution, determined by the component of $\nabla\times\boldsymbol{R}$ perpendicular to $\Bv$. The term “field line slippage rate” is, therefore, used here in a distinct sense, as a kinematic diagnostic of non-ideal field line behaviour rather than a transport velocity. Further} the slippage rate should also not be confused with the commonly used interpretation of the \emph{reconnection rate}, which is related to the rate of change of magnetic flux. {The slippage rate is not weighted by magnetic field strength and, therefore, does not directly measure the energetic importance of reconnection. Rather, it identifies where connectivity-changing non-ideal effects are active, while the associated energy release depends additionally on quantities, such as $\Sigv_D$.}

\section{Slippage rate as a proxy of reconnection in NLFFF extrapolations}
While the slippage rate may easily be calculated in simulations, we now consider how it may be applied to NLFFF extrapolations in order to provide a proxy for reconnection. First, NLFFF extrapolations only provide information about a magnetic field at a given time. The exact form of the resistivity is not prescribed and must be selected \emph{a posteriori} as a modelling choice. Our approach here is to consider a minimal model for the reconnection proxy and set the normalization $\eta\mu_0^{-1}=1$. This choice allows us to reveal directly how the NLFFF geometry is connected to the resistivity term without the need for additional modelling assumptions (particularly when information about $\eta$ may not be available). If more information about resistivity were available, the approach would be adaptable to account for this \citep[][]{MacTaggart2025Slippage}. 

{Given the above choice, the meaning of the slippage rate proxy for reconnection is the following: if the slippage rate is non-zero in part of the magnetic field at a given time, then the geometry contains sufficiently strong cross-field gradients to support connectivity-changing non-ideal evolution. We emphasize that, in the NLFFF framework, each extrapolation is treated as an independent snapshot of the magnetic field and the slippage rate provides information about reconnection at the given snapshot time.}

Further information can be found {about precisely how $\Sigv$ relates to the magnetic field geometry} through consideration of
the basic properties of NLFFFs themselves. First, the basic definition
of a NLFFF is that
\begin{equation}\label{nlfff_current}
    \nabla\times\Bv = \alpha\Bv,
\end{equation}
with 
\begin{equation}\label{alpha}
    \alpha =\frac{\Bv\cdot\nabla\times\Bv}{|\Bv|^2}.
\end{equation}
Equation (\ref{alpha}) measures the local twist of the magnetic field about a field line. In a NLFFF, $\alpha$ is constant on a given field line but has different values on different field lines. Using equation (\ref{nlfff_current}) with $\eta\mu_0^{-1}=1$, the resulting dissipation and slippage rates are 
\begin{equation}\label{force_free_rates}
    \Sigv_D =-\alpha^2\bv -\nabla\alpha\times\bv \quad {\rm and} \quad \Sigv = -\nabla\alpha\times\bv.
\end{equation}
In (\ref{force_free_rates}) there is a clear relationship between the NLFFF geometry and the dissipation and slippage rates. The dissipation rate contains terms parallel and perpendicular to $\bv$. The first of these describes the dissipation of the magnetic field along the direction of a field line. As per equation (\ref{Sigma_D}), this term represents pure (instantaneous) dissipation of the magnetic field strength and is not directly related to field line slippage (although dissipation may lead indirectly to greater cross-field gradients through magnetic field deformation). The second term of $\Sigv_D$ is the slippage rate $\Sigv$. This term shows that the field lines of a NLFFF subjected to resistivity will slip if there is a strong cross-field gradient in the local field-aligned twist. This relationship clearly reveals how the geometry of the NLFFF relates to the locations of reconnection. 

An important point to note here is that absolute slippage rates cannot be determined from extrapolations since only information about the magnetic field is available. However, the relative strength of the slippage rate in an extrapolation can be measured. This quantity is important as not only does the slippage rate say where reconnection will occur in a magnetic field, it will also indicate precisely where it is strongest. 

{The magnitude of the slippage rate defines a local kinematic timescale
\begin{equation}
\tau_{\Sigma} \sim |\Sigv|^{-1},
\end{equation}
which characterizes the rate at which a magnetic field line undergoes an $O(1)$ deviation from ideal evolution. Large values of $|\Sigv|$, therefore, correspond to rapid local changes in field line connectivity. In applications where the relevant physical scales are specified (e.g. time-dependent simulations), this timescale may be interpreted directly, whereas in NLFFF extrapolations it is, as mentioned above, used comparatively rather than absolutely.}

{Each extrapolation is treated as an independent snapshot, and the slippage rate provides a local diagnostic of reconnection activity within that configuration, rather than a measure of the cumulative reconnection required to evolve between successive states.}

\section{Relationship to QSLs}
The behaviour of reconnection in ARs has received considerable attention through the lens of QSL analysis \citep[][]{Demoulin2006QSLReview}. One particular success has been \emph{slip-running reconnection} \citep[][]{AulanierEtAl2006,DudikEtAl2014,LorincikEtAl2025}, whereby field lines may slip through QSLs to reconnect to distant parts of an AR - a purely 3D reconnection process. When we speak of ``a QSL,'' we refer to a region of limited volume within the magnetic field defined by a strong squashing factor $Q$. These regions are geometrically susceptible to reconnection, in the sense that even small non‑ideal effects can produce large changes in connectivity. In order to clarify how signatures of the slippage rate relate to strong squashing factor regions, we now derive a scaling estimate linking the magnitude of $\Sigv$ to the local properties of a QSL.

Let $D$ denote the Jacobian matrix of the magnetic field‑line mapping between two boundary surfaces. Following standard analysis, the squashing factor may be expressed as
\begin{equation}\label{squash_basic}
    Q=\frac{||D\|_F^2}{\det(D)},
\end{equation}
where $\|\cdot\|_F$ represents the Frobenius norm. If we denote $\sigma_+\ge\sigma_-> 0$ as the singular values of $D$, we may write
\begin{equation}
    \|D\|_F 
    ^2 = \sigma_+^2 + \sigma_-^2\quad {\rm and}\quad \det(D)=\sigma_+\sigma_-.
\end{equation}
In a strong QSL, the field line mapping is highly anisotropic, such that $\sigma_+\gg\sigma_-$, yielding the approximation
\begin{equation}
    Q \sim \frac{\sigma_+}{\sigma_-}.
\end{equation}
Provided that the mapping does not involve extreme area collapse or expansion, so that $\det(D)=\sigma_+\sigma_-\sim O(1)$, the dominant singular value scales as
\begin{equation}\label{sigma_Q_scale}
    \sigma_+\sim \sqrt{Q}.
\end{equation}
Geometrically, this corresponds to the stretching of an initially circular transverse neighbourhood of scale $\ell$ at one footpoint into an elongated structure of length $L\sim\sigma_+\ell$ at the conjugate footpoint \citep[see also][]{AulanierEtAl2006}. Hence,
\begin{equation}\label{l_Q_estimate}
    l\sim\frac{L}{\sqrt{Q}}.
\end{equation}
To relate this mapping scale to the slippage rate, we note that for a NLFFF, the slippage rate from equation (\ref{force_free_rates}) is controlled by cross-field gradients in $\alpha$. Assuming that $\alpha$ varies over the transverse scale $\ell$ within a QSL, we obtain the heuristic local scaling estimate
\begin{equation}\label{Sigma_Q_estimate}
    |\Sigv|\sim\frac{|\alpha|\sqrt{Q}}{L}.
\end{equation}
This scaling is intended as an order‑of‑magnitude estimate characterizing a typical slippage rate strength within a QSL rather than a point-wise equality.

Equation (\ref{Sigma_Q_estimate}) illustrates that strong squashing is a geometric amplifier of slippage only insofar as it creates small transverse length scales; it does not, by itself, guarantee significant reconnection. The presence of reconnection requires that these small scales coincide with appreciable gradients in the field‑aligned current, encoded by $\alpha$. In the limit $\alpha \to 0$, the slippage rate necessarily vanishes even for arbitrarily large $Q$. This reinforces the interpretation of QSLs as identifying regions of geometric susceptibility, while the slippage rate identifies regions where resistivity is physically active at a given time.

The slippage rate, therefore, complements QSL analysis by providing a local, instantaneous and physics‑weighted diagnostic of reconnection activity. Agreement between $Q$ and $|\Sigv|$ is expected in many flare‑productive ARs, but deviations between the two are themselves physically meaningful, reflecting variations in current structure and reconnection efficiency rather than shortcomings of either diagnostic.

\section{Application to AR11158}
We now apply the slippage and dissipation rates to NLFFF extrapolations of AR11158 with the aim of testing the theoretical considerations developed in Sections 4 and 5. Rather than simply demonstrating spatial coincidence with known reconnection proxies, we use the slippage rate to distinguish between magnetic configurations that are geometrically favourable for reconnection and those that exhibit signatures of active, resistivity‑induced field line slippage. AR11158 provides a particularly suitable test case, as its X2.2 flare has been studied extensively \citep[e.g.][to list but a few]{Schrijver2011ApJ,Tarr2013ApJ,Song2013RAA,Zhao2014ApJ,Kazachenko2015ApJ,Duan2025MNRAS} and its magnetic evolution is well constrained both prior to and following the eruptive event.



We make use of the NLFFF extrapolation algorithm by \cite{Jarolim2023NatAstron}, which has been shown to perform well in previous work when compared to other observations and metrics \citep[][]{Aslam2024MNRAS,MacTaggartWilliamsAslam2025}. This, of course, is just one choice of NLFFF extrapolation algorithm, and the following analysis is independent of the particular algorithm employed. For details of metrics related to the accuracy of the extrapolations of AR11158, we direct the reader to \cite{Aslam2024MNRAS}.

The extrapolations considered here range from 18:00 UT on 14/02/2011 to 02:48 on 15/02/2011 and cover the period of the X2.2 flare. The cadence is 12 minutes as the extrapolations are based on Space-Weather Helioseismic and Magnetic Imager (HMI) Active Region Patches (SHARP) vector magnetograms \citep[][]{Bobra2014SHARP}.

\subsection{General evolution}
The energization and onset of the X2.2 flare in AR11158 has been discussed in detail in the literature. Our focus here, therefore, is to concentrate on slippage rate signatures and their implications for reconnection.

\begin{figure*}
    \centering
    \begin{subfigure}[t]{0.48\textwidth}
        \centering
        \includegraphics[width=\linewidth]{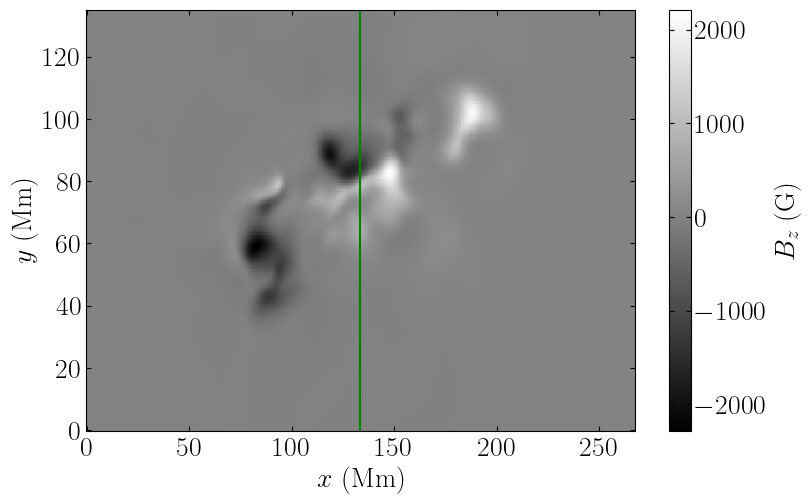}
        \caption{Magnetogram of $B_z$ at $z=0$ on 14/02/2011 at 18:00 UT. The green line indicates where $\max|\Sigv|$ and $\max|\Sigv_D|$ are calculated (at $z=2.88$ Mm). Note that this magnetogram has been smoothed compared to the original in order to make it applicable for a NLFFF extrapolation \citep[][]{Jarolim2023NatAstron}.}
        \label{fig:Bz_green_cut}
    \end{subfigure}
    \hfill
    \begin{subfigure}[t]{0.48\textwidth}
        \centering
        \includegraphics[width=\linewidth]{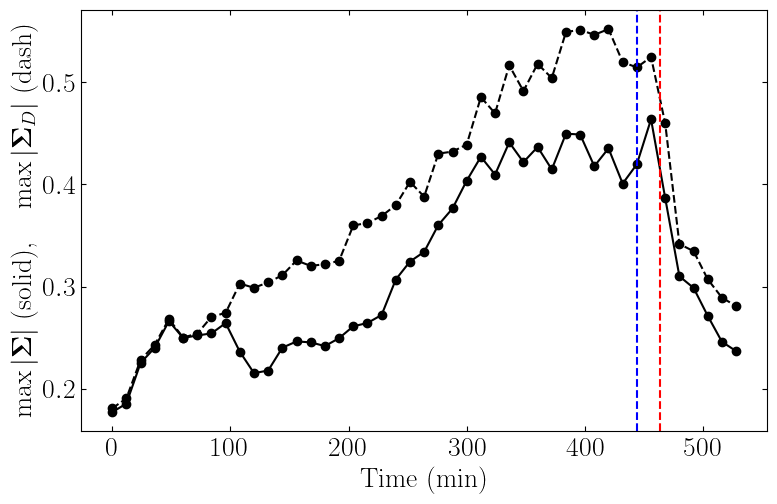}
        \caption{Time series of $\max|\Sigv|$ (solid) and $\max|\Sigv_D|$ (dash). The blue and red vertical lines indicate the times of the winding signature and the start of the X2.2 flare respectively. The start time $t=0$ corresponds to 14/02/2011 at 18:00 UT.}
        \label{fig:rates_in_time}
    \end{subfigure}

    \caption{A profile of the slippage and dissipation rates in the low corona across the polarity inversion line at $x=$ 133 Mm.}
    \label{fig:example}
\end{figure*}

The quadrupolar region contains a clearly identifiable polarity inversion line (PIL) at its centre. The polarities on either side are sheared and this is the source energy input into the magnetic field. Several studies have identified the presence of strong currents at the PIL \citep[e.g.][]{Wang2012AR11158,Sun2012AR11158,Wiegelmann2012AR11158,Janvier2013,Sun2015Currents,Duan2025MNRAS}, but what is the behaviour of reconnection in this location? To address this question, we take a cut through the PIL, as shown in Figure \ref{fig:Bz_green_cut} and examine the behaviour of the maxima of the dissipation and slippage rates in time. These quantities are shown in Figure \ref{fig:rates_in_time} and although the cut is taken across the entire length of the magnetogram, the maxima correspond to values at the PIL. The cut is placed near to the base of the extrapolation, at $z=2.88$ Mm, in order to focus on how the shear of the PIL relates to reconnection. 

Figure \ref{fig:rates_in_time} shows the temporal evolution of $\max|\Sigv_D|$ and $\max|\Sigv|$ measured along a cut through the main PIL at a height of 2.88 Mm. Initially, both quantities are comparable in magnitude. Over the subsequent $\sim$300 min, however, the two measures separate, with $\max|\Sigv_D|$ > $\max|\Sigv|$ and both increasing steadily.
This divergence indicates a change in the nature of non‑ideal processes in the low corona. The continued growth of $\max|\Sigv_D|$ reflects increasing field‑aligned dissipation associated with enhanced current concentrations along the PIL, while the simultaneous increase in $\max|\Sigv|$ demonstrates that a significant fraction of this dissipation is accompanied by cross‑field gradients in $\alpha$ and, therefore, by resistivity‑induced field‑line slippage. In other words, the non‑ideal evolution is not purely dissipative but involves changes in field line connectivity.

The maximum slippage rate reaches a plateau prior to flare onset, whereas the dissipation rate continues to grow for approximately an hour longer. This behaviour is consistent with a scenario in which the geometry of the magnetic field becomes increasingly favourable for reconnection up to a limiting configuration, after which further stress primarily contributes to field‑aligned dissipation rather than enhanced connectivity change. At around 450 min, both measures decrease rapidly after a transient growth, indicating a fundamental reconfiguration of the magnetic field associated with the flare and CME onset. This transition period is marked on Figure \ref{fig:rates_in_time} by blue and red dashed lines that indicate the times of the winding signature \citep[][]{Aslam2024MNRAS} and the start time of the X2.2 flare respectively. The winding signature is discussed below.

\begin{figure*}
\centering

\begin{subfigure}[t]{0.48\textwidth}
\centering
\begin{tikzpicture}
  \node[inner sep=0] (img)
    {\includegraphics[width=\linewidth]{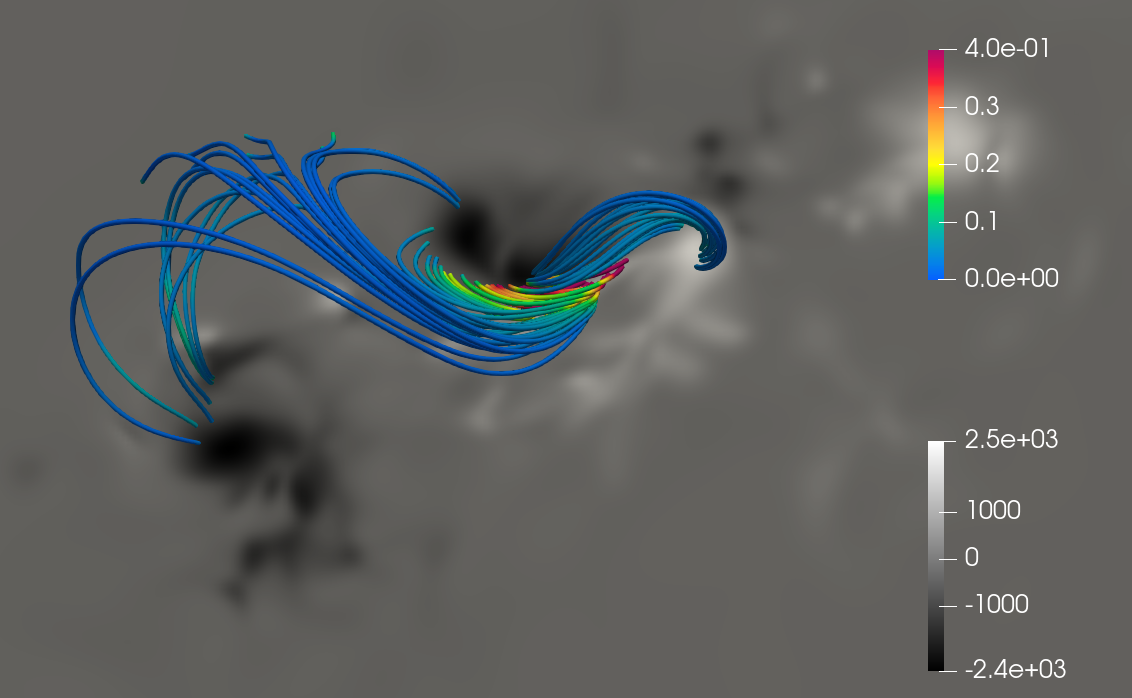}};


  \node[text=white, anchor=south east]
    at (4.2,1.2)
    {\small$|\Sigv_D|$};

  \node[text=white, anchor=south east]
    at (4.1,-1.8)
    {\small$B_z$};

  \draw[white,->,thick] (-3.5,-2) -- (-3,-2) node[right] {$x$};
  \draw[white,->,thick] (-3.5,-2) -- (-3.5,-1.5) node[above] {$y$};

\end{tikzpicture}

\caption{15/02/2011 at 01:00 UT}
\label{fig:dissipation_a}
\end{subfigure}
\hfill
\begin{subfigure}[t]{0.48\textwidth}
\centering
\begin{tikzpicture}
  \node[inner sep=0] (img)
    {\includegraphics[width=\linewidth]{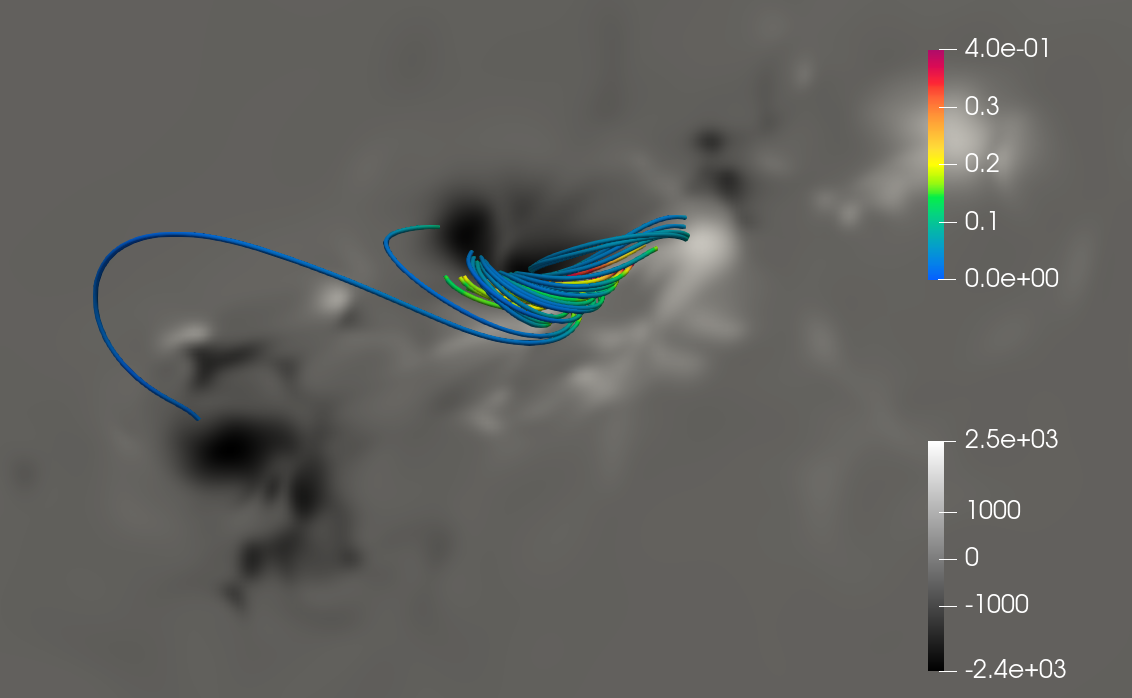}};

   \node[text=white, anchor=south east]
    at (4.2,1.2)
    {\small$|\Sigv_D|$};

  \node[text=white, anchor=south east]
    at (4.1,-1.8)
    {\small$B_z$};

  \draw[white,->,thick] (-3.5,-2) -- (-3,-2) node[right] {$x$};
  \draw[white,->,thick] (-3.5,-2) -- (-3.5,-1.5) node[above] {$y$};
\end{tikzpicture}

\caption{15/02/2011 at 02:48 UT}
\label{fig:dissipation_b}
\end{subfigure}

\caption{Field line extrapolations before and after the X2.2 flare. Field lines are traced from the same locations in (a) and (b), and are rendered with $|\Sigv_D|.$ The units of the line-of-sight magnetic field component $B_z$ are Gauss. {In relation to the timescale of Figure \ref{fig:rates_in_time}, (a) corresponds to 420 min and (b) to 528 min.}}
\label{fig:twocol}
\end{figure*}
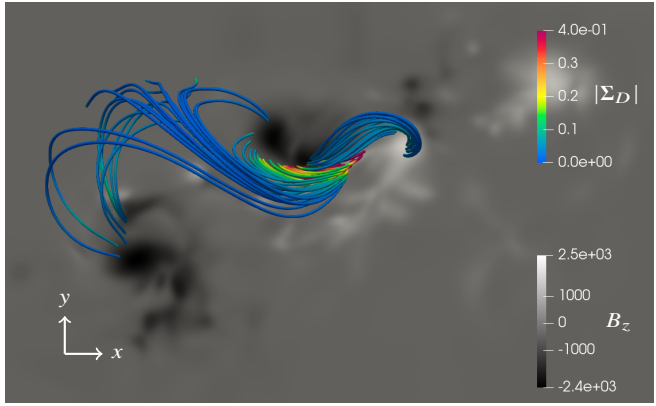
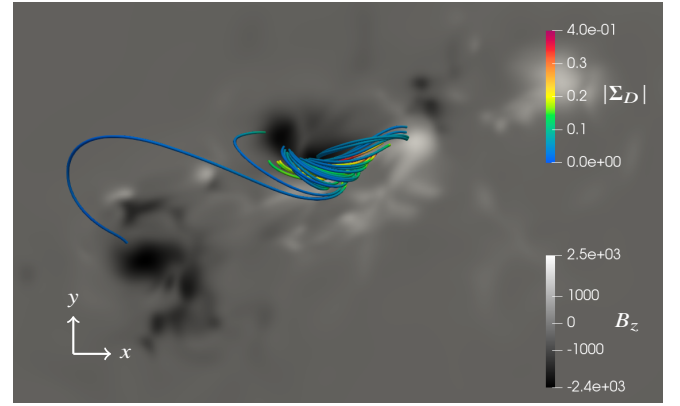

What the NLFFF extrapolations are capturing is a fundamental change in the magnetic field structure before and after the flare. The change in the magnetic field geometry influences the relative strengths and locations of the dissipation and slippage rates and, by implication, the behaviour of reconnection within the AR. To show this connection more clearly, field lines are traced before and after the flare in  Figures \ref{fig:dissipation_a} and \ref{fig:dissipation_b} respectively.

Field line extrapolations before and after the flare are shown in Figures \ref{fig:dissipation_a} and \ref{fig:dissipation_b}, using identical seed points centred on the PIL. Prior to the flare, the low‑lying field exhibits a strongly sheared ‘J‑shaped’ morphology \citep[][]{TitovDemoulin1999}, with field lines running nearly parallel to the PIL. These field lines coincide with regions of enhanced dissipation and slippage, indicating the presence of strong current gradients and active reconnection along an extended portion of the PIL.
Following the flare, the magnetic structure reorganizes into a less-sheared arcade whose field lines intersect the PIL at larger angles. In this configuration, both dissipation and slippage are significantly reduced, with enhanced values confined to a smaller subset of low‑lying field lines. This transition is consistent with flare‑driven relaxation and supports the interpretation that the observed reduction in the slippage and dissipation rates reflects the cessation of large‑scale reconnection rather than merely a redistribution of currents.

\subsection{Identifying the winding signature}
It was shown by \cite{Aslam2024MNRAS} that a sharp and transient increase in the magnetic winding flux measured at the photosphere is a common precursor to CME onset. Such winding signatures have been interpreted as indicators of a rapid reorganisation of the coronal magnetic field, signalling the approach to an eruption. Importantly, these signatures do not necessarily coincide with the main PIL, but may occur at locations where the coronal field geometry undergoes a qualitative transition. AR11158 provides a clear example of this behaviour.

\begin{figure*}
\centering

\begin{subfigure}[t]{0.48\textwidth}
\centering
\begin{tikzpicture}
  \node[inner sep=0] (img)
    {\includegraphics[width=\linewidth]{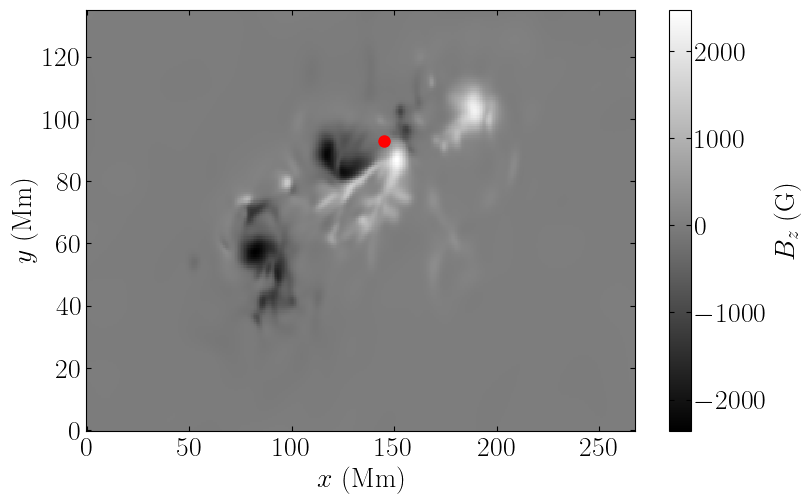}};





\end{tikzpicture}

\caption{Line-of-sight magnetogram with the location of the winding signature indicated by a red circle.}
\label{fig:wind_a}
\end{subfigure}
\hfill
\begin{subfigure}[t]{0.48\textwidth}
\centering
\begin{tikzpicture}
  \node[inner sep=0] (img)
    {\includegraphics[width=\linewidth]{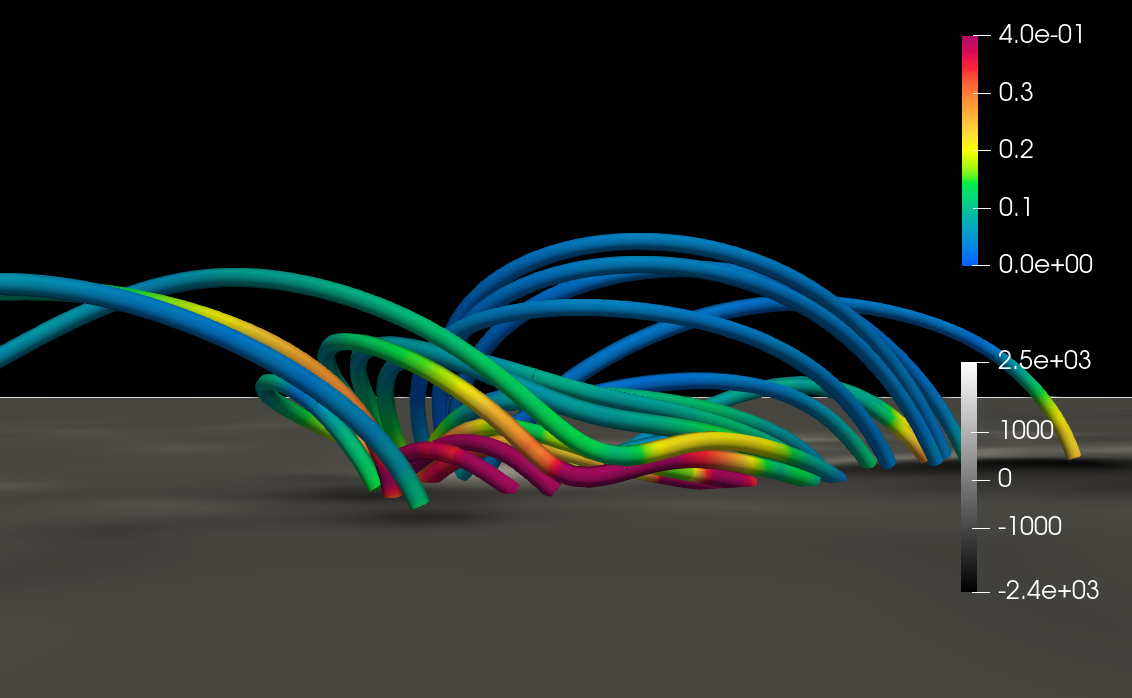}};

   \node[text=white, anchor=south east]
    at (4.2,1.3)
    {\small$|\Sigv|$};

  \node[text=white, anchor=south east]
    at (4.15,-1.2)
    {\small$B_z$};

  \draw[white,->,thick] 
  (-3,-2) -- (-3.5,-2) node[left] {$x$};
  \draw[white,->,thick] (-3,-2) -- (-3,-1.5) node[above] {$z$};
\end{tikzpicture}

\caption{Field line structure at the location of the winding signature. Field lines are rendered with $|\Sigv|$ and $B_z$ is measured in Gauss.}
\label{fig:wind_b}
\end{subfigure}

\caption{Details relating to the location and structure of the magnetic winding signature on 15/02/2011 at 02:48 UT.}
\label{fig:twocol}
\end{figure*}
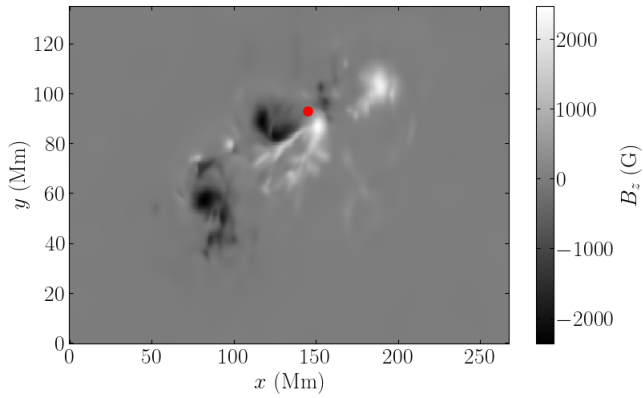
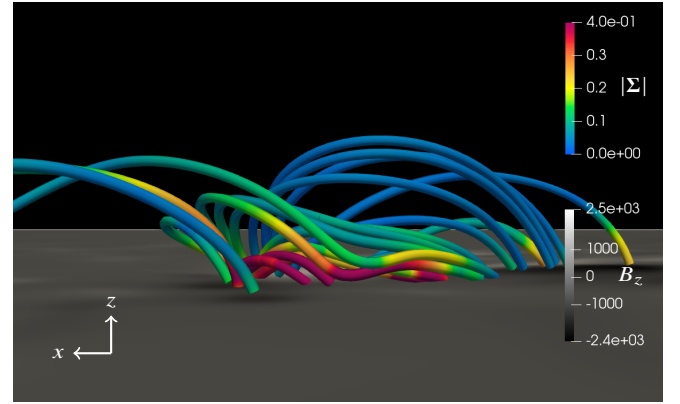
A pronounced winding signature was identified in AR11158 by \cite{Aslam2024MNRAS} on 15/02/2011 at 01:24 UT, at the photospheric location marked by the red circle in Figure \ref{fig:wind_a}. This location lies away from the main PIL and is, therefore, not associated with the strongest photospheric shear or current concentrations. As such, it provides a useful test of whether the slippage rate is capable of identifying reconnection-related structure beyond the regions highlighted by conventional PIL-based diagnostics.

The magnetic structure associated with the winding signature is shown in Figure \ref{fig:wind_b}, where the field is viewed from the negative $y$-direction. The field lines forming the upper branch of the characteristic J‑shaped structure (cf. Figure \ref{fig:dissipation_a}) exhibit a pronounced dip in this region and intersect the photosphere tangentially. This configuration corresponds to a bald patch, where magnetic field lines graze the photosphere and the normal component of the magnetic field changes sign along the field line \citep[][]{TitovPriestDemoulin1993}. Bald patches are well-established sites of reconnection, particularly in strongly sheared magnetic configurations \citep[][]{PontinPriest2022}.

Consistent with this interpretation, the slippage rate $|\Sigv|$ is strongly enhanced in the vicinity of the winding signature. { This behaviour shows that the winding signature coincides with an enhanced slippage rate, consistent with its interpretation as a reconnection-related feature of the magnetic field, rather than merely a region of geometrically complex field line mapping.}

The slippage rate enhancement at the winding signature location is also temporally localized. It appears shortly before the onset of the X2.2 flare and diminishes rapidly during the post‑flare relaxation phase, mirroring the behaviour of the winding flux itself. This timing further supports the interpretation that the winding signature marks a critical change in the coronal field connectivity, associated with the initiation of large-scale reconnection during the eruption.

\subsection{Flare ribbons}
One of the most prominent observational signatures of magnetic reconnection associated with flares and CMEs is the appearance of \emph{flare ribbons} in the lower solar atmosphere \citep[][]{Fletcher2011FlareReview}. These elongated regions of enhanced emission, typically observed in UV and EUV, trace the chromospheric and upper‑photospheric footpoints of newly reconnected coronal field lines. As such, flare ribbons provide an important observational bridge between coronal magnetic structure and reconnection-driven energy release, and provide an important constraint on magnetic field models of ARs.

The flare ribbons of AR11158 have been studied extensively using QSL analysis, where their locations have been shown to coincide closely with a hyperbolic flux tube (HFT) configuration implicated in large-scale reconnection during the eruption \citep[][]{Zhao2014ApJ}. This system, therefore, provides a well-established benchmark against which the slippage rate proxy of reconnection may be tested.

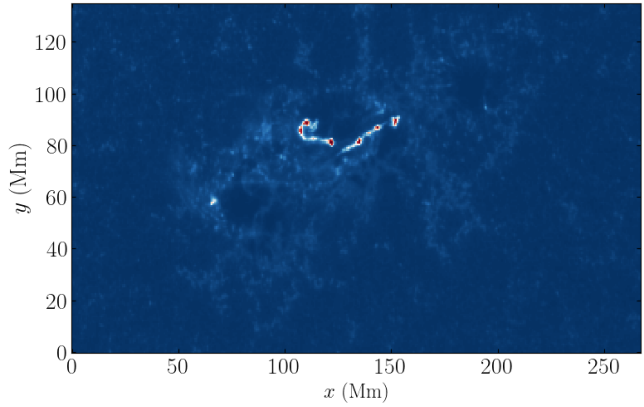
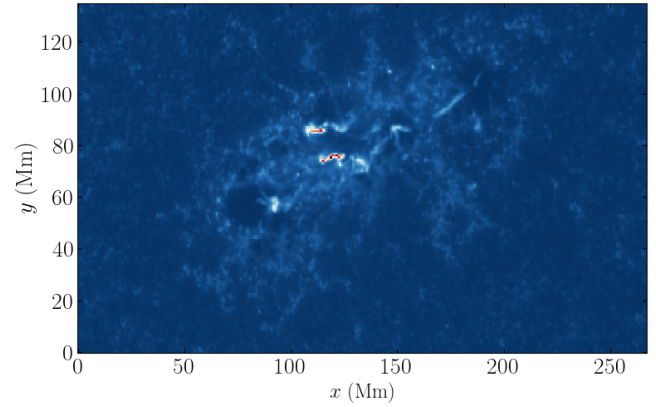
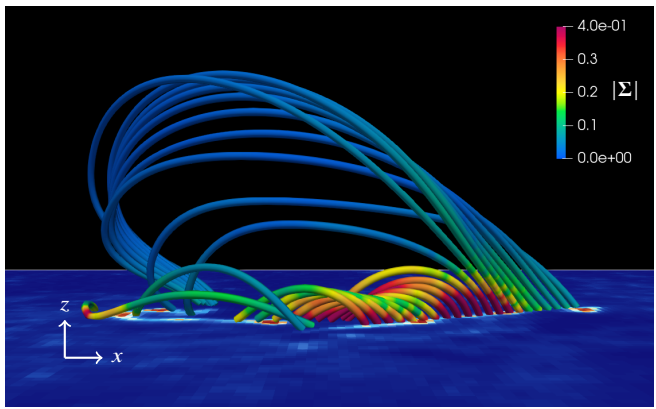
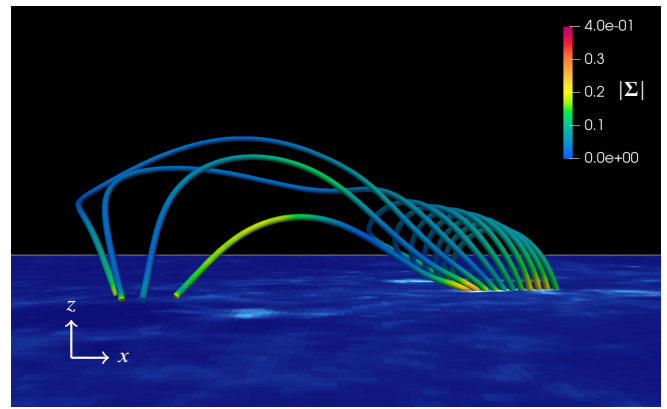
\begin{figure*}
\centering

\begin{subfigure}[t]{0.48\textwidth}
\centering
\begin{tikzpicture}
  \node[inner sep=0] (img)
    {\includegraphics[width=\linewidth]{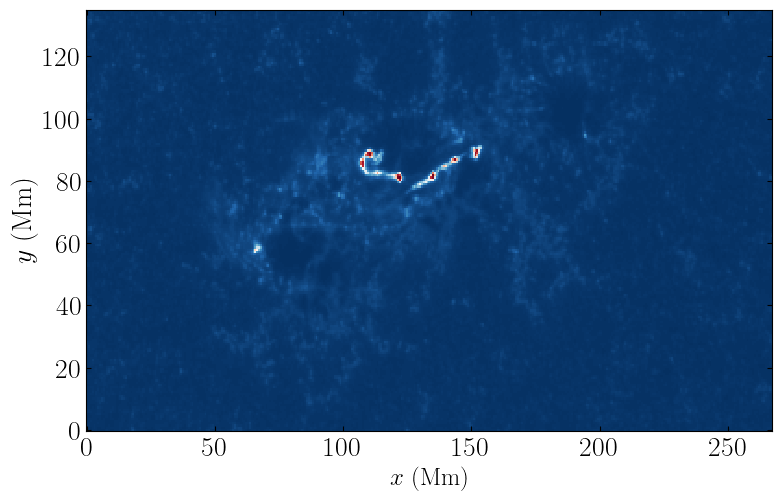}};



\end{tikzpicture}

\caption{AIA 1600 \r{A} on 15/02/2011 at 01:47 UT}
\label{fig:aia_a}
\end{subfigure}
\hfill
\begin{subfigure}[t]{0.48\textwidth}
\centering
\begin{tikzpicture}
  \node[inner sep=0] (img)
    {\includegraphics[width=\linewidth]{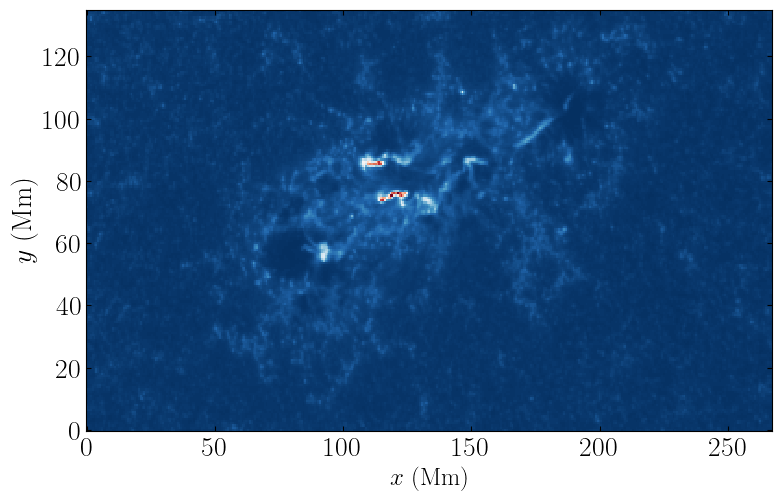}};


\end{tikzpicture}
\caption{AIA 1600 \r{A} on 15/02/2011 at 02:47 UT}
\label{fig:aia_b}
\end{subfigure}

\vspace{1cm}

\begin{subfigure}[t]{0.48\textwidth}
\centering
\begin{tikzpicture}
  \node[inner sep=0] (img)
    {\includegraphics[width=\linewidth]{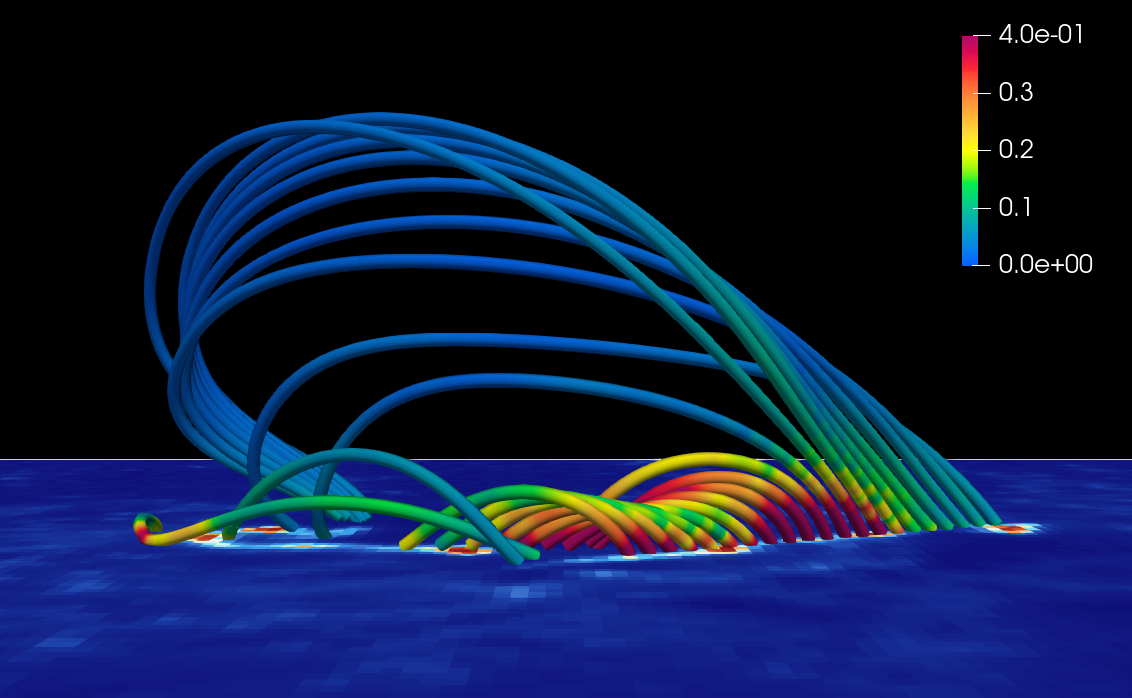}};


  \node[text=white, anchor=south east]
    at (4.2,1.3)
    {\small$|\Sigv|$};

\draw[white,->,thick] (-3.5,-2) -- (-3,-2) node[right] {$x$};
  \draw[white,->,thick] (-3.5,-2) -- (-3.5,-1.5) node[above] {$z$};
  
\end{tikzpicture}

\caption{NLFFF extrapolation on 15/02/2011 at 01:48 UT}
\label{fig:aia_c}
\end{subfigure}
\hfill
\begin{subfigure}[t]{0.48\textwidth}
\centering
\begin{tikzpicture}
  \node[inner sep=0] (img)
    {\includegraphics[width=\linewidth]{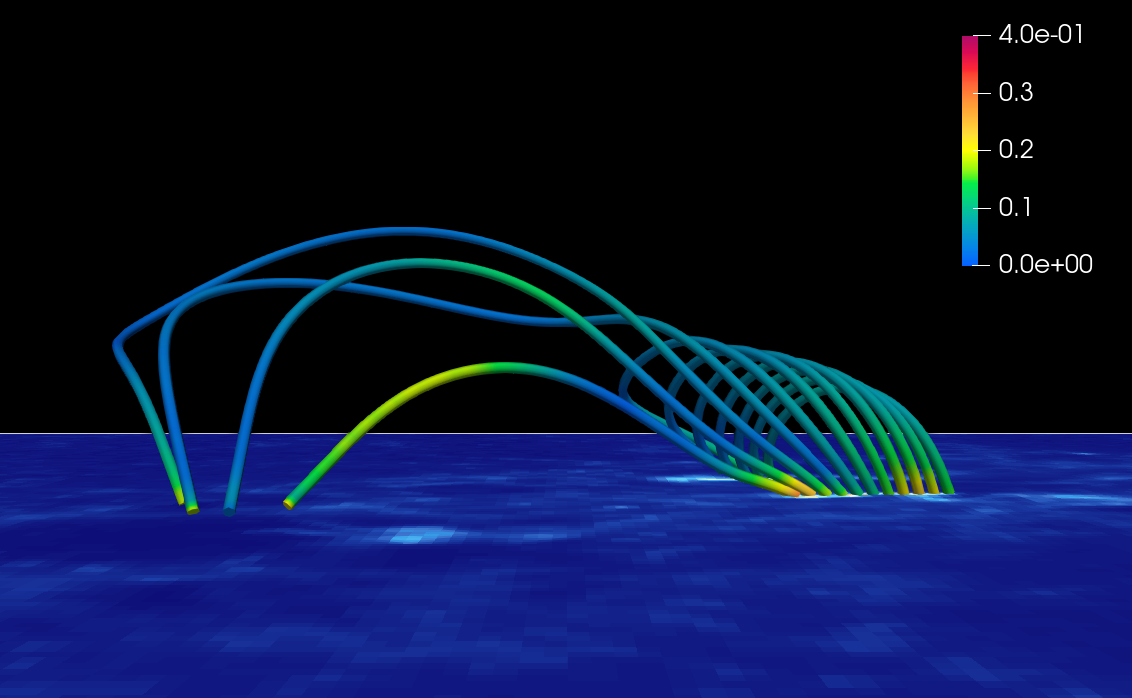}};

   \node[text=white, anchor=south east]
    at (4.2,1.3)
    {\small$|\Sigv|$};

\draw[white,->,thick] (-3.5,-2) -- (-3,-2) node[right] {$x$};
  \draw[white,->,thick] (-3.5,-2) -- (-3.5,-1.5) node[above] {$z$};

\end{tikzpicture}
\caption{NLFFF extrapolation on 15/02/2011 at 02:48 UT}
\label{fig:aia_d}
\end{subfigure}

\caption{Maps of AIA 1600 \r{A} reveal the flare ribbons; (a) near the start of the flare and (b) in the post-flare relaxation phase. Corresponding field line extrapolations, traced from the ribbon locations at $z=0$ are shown in (c) and (d). Field lines are rendered with $|\Sigv|$. There is a 1-minute difference between AIA and HMI data.}
\label{fig:twocol}
\end{figure*}

Flare ribbons are readily identified using 1600 \r{A} observations from the Atmospheric Imaging Assembly \citep[AIA;][]{Lemen2012AIA}. Figures \ref{fig:aia_a} and \ref{fig:aia_b} show AIA 1600 \r{A} images capturing the ribbon morphology during the early impulsive phase and the post‑flare phase, respectively. 

To relate these observations to the coronal magnetic field, as modelled by the NLFFF extrapolations, Figure \ref{fig:aia_c} displays field lines traced from locations along the right‑hand ribbon in Figure \ref{fig:aia_a} and rendered with the magnitude of the slippage rate. The strongest slippage rates are found along field lines whose footpoints coincide with the observed ribbon emission. This close correspondence demonstrates that the flare ribbons are not only associated with regions of large connectivity gradients, but with locations of active resistivity-induced field line slippage in the extrapolated coronal field. In other words, {the slippage rate highlights that these ribbon footpoints coincide with enhanced field line slippage, consistent with independently inferred reconnection activity}.

Importantly, this agreement goes beyond the geometric association established by QSL analysis alone. While QSLs identify where reconnection may occur based on topology and mapping properties, the slippage rate identifies where reconnection is instantaneously active within the magnetic field. The present results, therefore, provide a direct physical interpretation of the ribbon–QSL correspondence reported by \cite{Janvier2014Currents}. 

The NLFFF extrapolation also reveals strongly twisted and dipped field lines associated with the hooked portion of the left‑hand ribbon in Figure \ref{fig:aia_a}. The field line connecting to this hooked region in Figure \ref{fig:aia_c}, which is traced from the other ribbon, exhibits a strong slippage rate. This result is consistent with interpretations of hooked ribbons as signatures of reconnection involving the erupting flux rope and its surrounding magnetic environment \citep[][]{Janvier2014Currents}. 

Figure \ref{fig:aia_b} shows the ribbon morphology during the post‑flare phase, where the emission has reorganized into two near‑parallel ribbons characteristic of a post‑flare arcade. Field lines traced from the lower ribbon and rendered with $|\Sigv|$ in Figure \ref{fig:aia_d} form a comparatively less‑sheared arcade structure (with the exception of some field lines connecting to distant parts of the AR) relative to the earlier phase. The overall reduction in $|\Sigv|$ reflects a substantial decrease in resistivity‑induced connectivity change, consistent with the relaxation of the magnetic field.

\subsection{Comparison to the squashing factor}
In this section we examine in detail the correspondence between the field line slippage rate magnitude $|\Sigv|$ and the squashing factor $Q$, using the NLFFF extrapolation of AR11158, shortly before flare onset, as a representative example. The purpose of this comparison is not simply to assess spatial overlap between the two quantities, but to clarify what additional physical information is provided by the slippage rate beyond that already available from QSL analysis.

\begin{figure*}
\centering

\begin{subfigure}[t]{0.48\textwidth}
\centering
\begin{tikzpicture}
  \node[inner sep=0] (img)
    {\includegraphics[width=\linewidth]{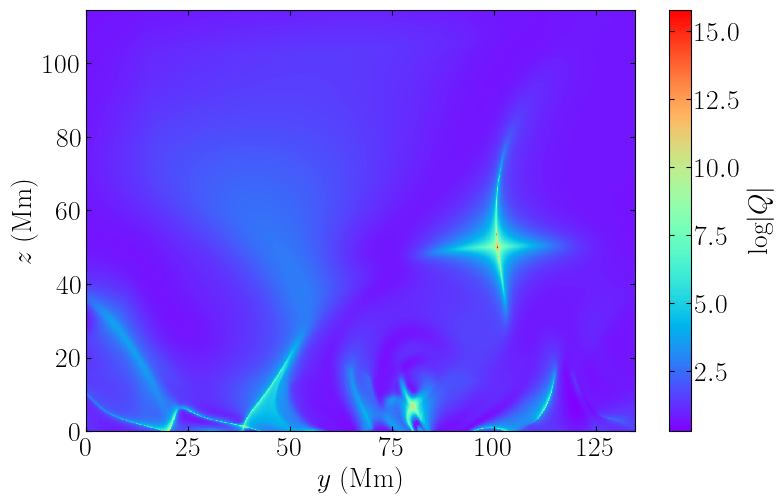}};



\end{tikzpicture}

\caption{Map of $\log|Q|$.}
\label{fig:qsl_a}
\end{subfigure}
\hfill
\begin{subfigure}[t]{0.48\textwidth}
\centering
\begin{tikzpicture}
  \node[inner sep=0] (img)
    {\includegraphics[width=\linewidth]{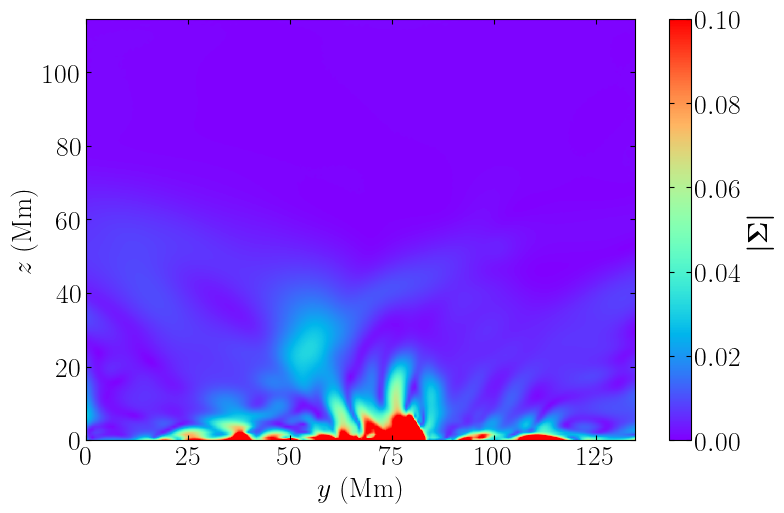}};


\end{tikzpicture}
\caption{Map of $|\Sigv|.$}
\label{fig:qsl_b}
\end{subfigure}

\vspace{1cm}

\begin{subfigure}[t]{0.48\textwidth}
\centering
\begin{tikzpicture}
  \node[inner sep=0] (img)
    {\includegraphics[width=\linewidth]{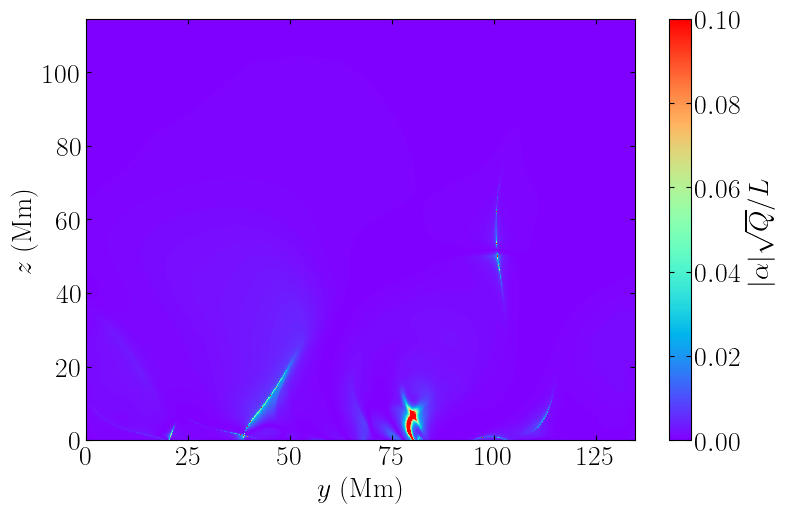}};


\end{tikzpicture}

\caption{Map of the estimate in (\ref{Sigma_Q_estimate}).}
\label{fig:qsl_c}
\end{subfigure}
\hfill
\begin{subfigure}[t]{0.48\textwidth}
\centering
\begin{tikzpicture}
  \node[inner sep=0] (img)
    {\includegraphics[width=\linewidth]{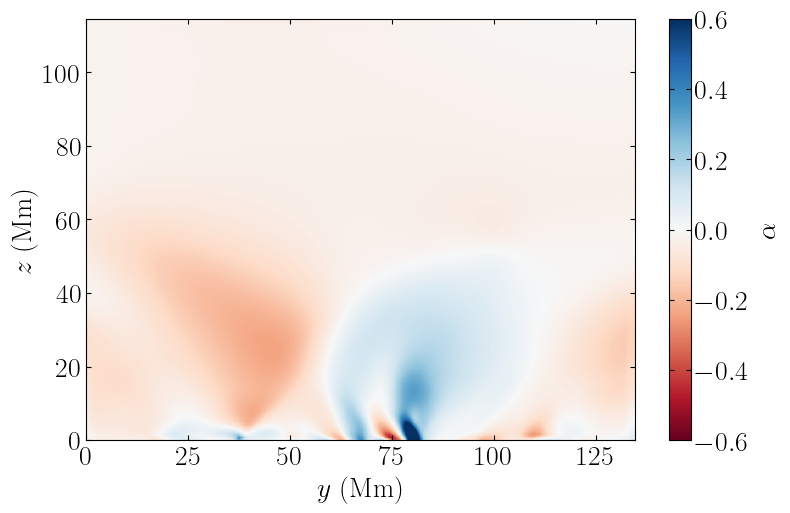}};

  
\end{tikzpicture}
\caption{Map of $\alpha$.}
\label{fig:qsl_d}
\end{subfigure}

\caption{Maps in the $(y,z)$-plane of quantities related to reconnection proxies. Slices are taken at $x=130$ Mm on 15/02/2011 at 01:36 UT. The resolution for the squashing factor calculation is 1000$^2$.}
\label{fig:twocol}
\end{figure*}

We focus on the extrapolated field at 01:36 UT, shortly before the X2.2 flare, and consider a vertical slice across the main PIL at $x=130$ Mm. Figure \ref{fig:qsl_a} shows a map of $\log|Q|$ calculated using \texttt{UFiT} \citep[][]{Aslanyan2024UFiT}, with a field line mapping resolution of 1000$^2$. Several high‑$Q$ features are apparent in this slice. In particular, two regions exhibit especially strong squashing: (i) a low-lying structure at the PIL corresponding to strongly-sheared field, and (ii) a null point located higher in the corona. The geometrical measure of the squashing factor identifies these locations as the most favourable for reconnection.

Figure \ref{fig:qsl_b} shows the corresponding map of the slippage rate magnitude for the same slice. For clarity, the colour scale has been capped at $|\Sigv| = 0.1$ to highlight the spatial structure of the signature rather than its absolute extrema. A key result is immediately evident: the strongest slippage rate signal is concentrated in the low corona at the PIL, coincident with the sheared-field QSL, whereas no comparably strong slippage rate signal is associated with the null point identified in the squashing factor map.

This partial correspondence — strong agreement at the PIL but a clear discrepancy at the null point — is central to the present analysis. It demonstrates that large squashing factors identify regions of geometric susceptibility to reconnection, but do not by themselves determine where reconnection is physically active in the extrapolated field. In contrast, the slippage rate selectively highlights those QSLs where strong cross‑field gradients in the field‑aligned twist parameter $\alpha$ are present, and, therefore, where resistivity‑induced field line slippage is expected to occur within the NLFFF approximation.

The absence of a strong slippage rate signature at the null point implies that, at 01:36 UT, the magnetic field in the vicinity of the null is close to being locally linear force‑free or potential. In such a configuration, the gradients of $\alpha$ are weak, and thus the slippage rate, which depends on cross-field gradients in $\alpha$, remains small. The correct interpretation of Figure \ref{fig:qsl_b} is, therefore, not that reconnection cannot occur at the null point, but that this particular null point at that particular time is not a dynamically significant reconnection site, despite its strong geometric signature in $Q$.

To further explore the relationship between $|\Sigv|$ and $Q$, Figure \ref{fig:qsl_c} shows a map of the $Q$-based estimate of the slippage rate magnitude (\ref{Sigma_Q_estimate}). Here we take $L=150$ Mm as an estimate of the characteristic field‑line length in the active region. Moderately reducing this value (e.g. to $L=100$ Mm) does not alter the qualitative conclusions. The estimated map captures the dominant slippage structure at the PIL but suppresses most other features seen in the squashing factor map. In particular, the signature associated with the null point is substantially weakened relative to that in $\log|Q|$. 

Residual weak features near the null point in Figure \ref{fig:qsl_c} continue to diminish as the resolution of the $Q$-map is increased up to 1000$^2$, indicating that they arise from the numerical amplification of small-scale mapping gradients rather than from physically meaningful current‑structure variations. This behaviour reinforces the interpretation that strong $Q$ alone does not guarantee significant slippage, and that the addition of current‑related information through $\alpha$ is essential for identifying physically active reconnection sites.

Finally, Figure \ref{fig:qsl_d} shows the distribution of $\alpha$ in the same slice. This map confirms that the region surrounding the null point is close to potential, whereas strong cross‑field gradients in $\alpha$ are present in the low corona near the PIL, precisely where the slippage rate attains its maximum values. Together, Figures \ref{fig:qsl_b} to \ref{fig:qsl_d} demonstrate explicitly how the slippage rate refines QSL analysis by weighting geometric complexity with physically relevant current structure.

Taken together, these results illustrate that the slippage rate should be viewed as a physics‑weighted refinement of QSL diagnostics rather than as a replacement for them. QSLs identify where reconnection may occur based on magnetic connectivity, while the slippage rate identifies where reconnection is instantaneously active within a given magnetic configuration. Disagreements between $Q$ and $|\Sigv|$, such as the example presented here, are, therefore, not deficiencies of either diagnostic but instead provide valuable insight into the evolving reconnection landscape of the AR.

\section{Conclusions}

In this paper, we have investigated the application of the field line slippage rate as a proxy for three‑dimensional magnetic reconnection in nonlinear force‑free field (NLFFF) extrapolations, with the aim of connecting the geometry of the coronal magnetic field directly to the non‑ideal terms in Ohm’s law. Unlike purely geometrical or topological diagnostics, the slippage rate provides a local and instantaneous diagnostic of (for the application considered here) resistivity‑induced deviations from ideal field line motion, enabling a physically grounded interpretation of where and how reconnection is occurring within an extrapolated magnetic field.

We first clarify the theoretical meaning of the slippage rate and its relationship to reconnection. For NLFFFs, and adopting a minimal resistive normalization, we show explicitly that the slippage rate is controlled by cross‑field gradients of the field‑aligned twist $\alpha$. This result establishes a direct link between current structure and reconnection activity, and demonstrates that reconnection signatures are governed not only by magnetic connectivity but also by how current is distributed across neighbouring field lines.

We then examine the relationship between the slippage rate and quasi‑separatrix layers (QSLs). By deriving a simple scaling estimate relating $|\Sigv|$ to the squashing factor $Q$, we show that strong squashing acts as a geometric amplifier of slippage only insofar as it produces small transverse length scales. Large values of $Q$ alone do not guarantee reconnection: a significant slippage rate requires that these geometric features coincide with substantial cross‑field gradients in $\alpha$. This distinction clarifies the complementary roles of QSL analysis and slippage rate‑based diagnostics and provides a principled framework for interpreting agreements and discrepancies between them.

Applying this framework to a sequence of NLFFF extrapolations of AR11158, we demonstrate that the slippage rate reveals clear signatures of enhanced three‑dimensional reconnection in the low corona during the build‑up to the X2.2 flare. Along the main polarity inversion line (PIL), the temporal evolution of the slippage rate shows that non‑ideal evolution prior to the flare is not purely dissipative but involves genuine resistivity‑induced field line slippage. The subsequent reduction of $|\Sigv|$ after flare onset is consistent with a relaxation of the coronal magnetic field toward a less sheared configuration.

We further show that the slippage rate {highlights locations consistent with reconnection activity that are} associated with magnetic winding signatures away from the main PIL, including bald patch geometries linked to the evolving flux rope structure. These results demonstrate that $\Sigv$ is sensitive to reconnection in strongly three‑dimensional configurations.

Finally, by comparing maps of $|\Sigv|$ and $Q$ directly, we demonstrate how the slippage rate refines QSL analysis in practice. While strong squashing at the PIL corresponds to strong slippage, other high‑$Q$ features, such as a coronal null point identified in the same extrapolation, exhibit little or no slippage rate signal, reflecting weak cross‑field gradients in $\alpha$. These differences are physically meaningful: they indicate that not all QSLs or topological features are dynamically important reconnection sites at a given time. The slippage rate,  therefore, provides a physics‑weighted diagnostic that distinguishes between regions of geometric susceptibility and regions of active reconnection.

Overall, this work demonstrates that the field line slippage rate is a powerful and complementary tool for the analysis of magnetic reconnection in NLFFF extrapolations. When used alongside established topological and quasi‑topological diagnostics such as QSLs, it allows for a more complete and physically informed picture of where three‑dimensional reconnection is occurring and how its role evolves during eruptive events. Future work will explore the application of the slippage rate to larger samples of eruptive and non‑eruptive active regions, as well as its extension to time‑dependent and data‑driven coronal models.

\section*{Acknowledgements}
We are very grateful for insightful discussions with Lyndsay Fletcher. S.S. acknowledges support from a Maclaurin Scholarship from the University of Glasgow. D.M. acknowledges support from a Leverhulme Trust grant (RPG-2023-182), a  Science and Technologies Facilities Council (STFC) grant (ST/Y001672/1) and a Personal Fellowship from the Royal Society of Edinburgh (ID: 4282). Results were obtained using the ARCHIE-WeSt High Performance Computer (www.archie-west.ac.uk) based at the University of Strathclyde.

\section*{Data Availability}

The \texttt{NF2} code for downloading SHARP data and creating NLFFF extrapolations can be downloaded from: 

\texttt{https://github.com/RobertJaro/NF2/tree/main}.

\vspace{0.2cm}

\noindent The \texttt{UFiT} code for calculating the squashing factor maps can be downloaded from: 

\texttt{https://github.com/Valentin-Aslanyan/UFiT}.



\bibliographystyle{mnras}
\bibliography{slippage} 





\bsp	
\label{lastpage}
\end{document}